\documentclass[%
 reprint,
superscriptaddress,
 aps,
 prl, 
]{revtex4-1}

\usepackage{graphicx}
\usepackage{dcolumn}
\usepackage{bm}
\usepackage{hyperref}

\usepackage[version=3]{mhchem}

\usepackage[range-units = single,range-phrase = \text{--},separate-uncertainty = true,detect-all]{siunitx}
	\DeclareSIUnit\linepair{lp}
	\DeclareSIUnit\pixels{px}

\begin{document}

\title{Mean path length invariance in multiple light scattering}

\author{Romolo Savo}
 \affiliation{%
Laboratoire Kastler Brossel, UMR 8552, CNRS, Ecole Normale Sup\'{e}rieure, Universit\'{e} Pierre et Marie Curie, Coll\`{e}ge de France, 24 rue Lhomond, 75005 Paris, France
}%

\author{Romain Pierrat}
\affiliation{
 ESPCI Paris, PSL Research University, CNRS, Institut Langevin, 1 rue Jussieu, 75005, Paris, France
}%

\author{Ulysse Najar}
\affiliation{%
Laboratoire Kastler Brossel, UMR 8552, CNRS, Ecole Normale Sup\'{e}rieure, Universit\'{e} Pierre et Marie Curie, Coll\`{e}ge de France, 24 rue Lhomond, 75005 Paris, France
}%

\author{R\'emi Carminati}
\affiliation{
 ESPCI Paris, PSL Research University, CNRS, Institut Langevin, 1 rue Jussieu, 75005, Paris, France
}%

\author{Stefan Rotter}
\affiliation{
 Institute for Theoretical Physics, Vienna University of Technology (TU Wien), Vienna, A-1040, Austria
}%

\author{Sylvain Gigan}
 \affiliation{%
Laboratoire Kastler Brossel, UMR 8552, CNRS, Ecole Normale Sup\'{e}rieure, Universit\'{e} Pierre et Marie Curie, Coll\`{e}ge de France, 24 rue Lhomond, 75005 Paris, France
}%

\date{\today}

\begin{abstract}
Our everyday experience teaches us that the structure of a medium strongly influences how light propagates through it. A disordered medium, e.g., appears transparent or opaque, depending on whether its structure features a mean free path that is larger or smaller than the medium thickness. While the microstructure of the medium uniquely determines the shape of all penetrating light paths, recent theoretical insights indicate that the mean length of these paths is entirely independent of any structural medium property and thus also invariant with respect to a change in the mean free path. Here, we report an  experiment that demonstrates this surprising property explicitly. Using colloidal solutions with varying concentration and particle size, we establish an invariance of the mean path length spanning nearly two orders of magnitude in scattering strength, from almost transparent to very opaque media. This very general, fundamental and counterintuitive result can be extended to a wide range of systems, however ordered, correlated or disordered, and has important consequences for many fields, including light trapping and harvesting for solar cells and more generally in photonic structure design.
\end{abstract}

\maketitle

\textit{General introduction.}
Wave transport in complex media is at the heart of many disciplines (optics, acoustics, electronics, quantum physics),
and encompasses a wide variety of situations from the micro to the macro scale~\cite{sheng2006introduction}. The associated transport phenomena
comprise a wealth of applications in microelectronics, photonics, medicine, atmospheric science,
soft matter, lasers, solar cells, photonic crystals, and bioimaging, to cite just a few~\cite{joannopoulos1997photonic,ntziachristos2010going,wiersma2008randomlasers,lu2014topologicalphotonics,polman2012photonic,Rojas-Ochoa2004photonicolloids,marshak20053d}. In all these domains, the
structure of the complex medium is inherently linked to its physical behavior, in particular to the properties of waves
scattering through this medium. Correspondingly, much of the progress in these fields has been linked to the ability to
modify and engineer the medium structure such as to fulfill a desired purpose~\cite{blanco2000opals,hu2008localization, Cao2009randomlaser,barthelemy2008levy,vynck2012photonmenagement,florescu2009designer,riboli2014engineeringlightconf,Muskens:2008kn,Garnett:2010en}. 

\begin{figure*}[t!]
   \centering
   \includegraphics[scale=1]{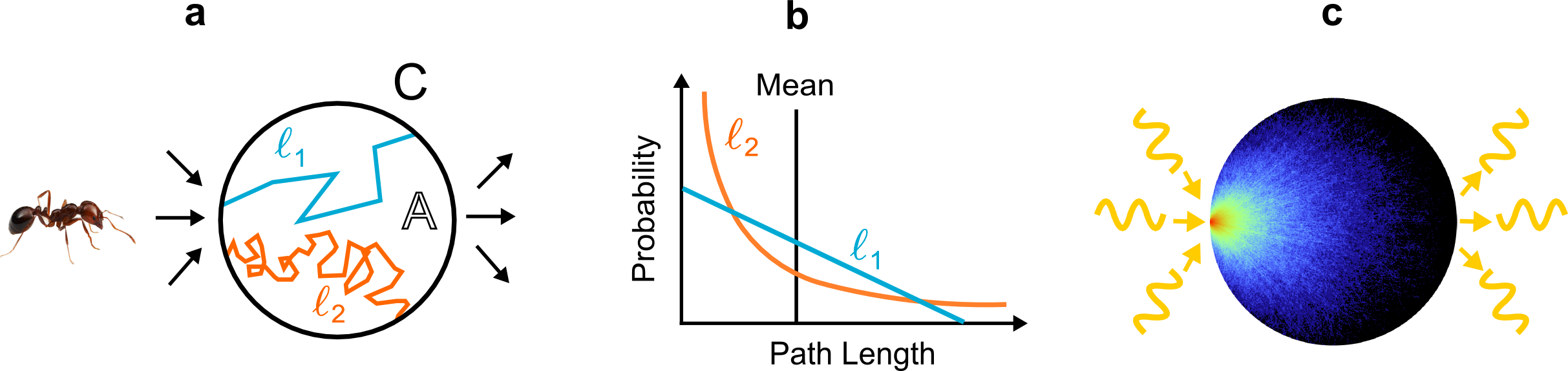}
   \caption{Random walks are ubiquitous in nature, they
   describe the erratic motion of ants just as well as the propagation of waves in disordered scattering
   media.~\textbf{a}, For random walks the mean length $\langle s \rangle$ of trajectories crossing a bounded region
   does not depend on the characteristic mean free path, $\ell$ , but only on the ratio between the surface $A$ and the
   perimeter $C$, i.e., $\langle s \rangle=\pi A/C$ (in two dimensions). Here we represent two trajectories, corresponding to two different mean free path $\ell_1 \gg \ell_2$. For this fundamental result to hold the mean value
   $\langle \ldots \rangle$ needs to include all surface positions for entering the medium and an isotropic distribution
   of incidence angles.~\textbf{b}, Illustration of the variation in the path length distribution for the two mean free
   paths $\ell_1$ and $\ell_2$ illustrated in (\textbf{a}). When changing the mean free path, the path length
   distribution changes dramatically, but its mean values remains the same.~\textbf{c}, The same physics also
   applies to light scattering through a disordered region - as illustrated here for light rays entering a circular
   region at a specific point with isotropic illumination.}
   \label{fig:figure1}
\end{figure*}

\begin{figure*}[t!]
   \centering
    \includegraphics[scale=1]{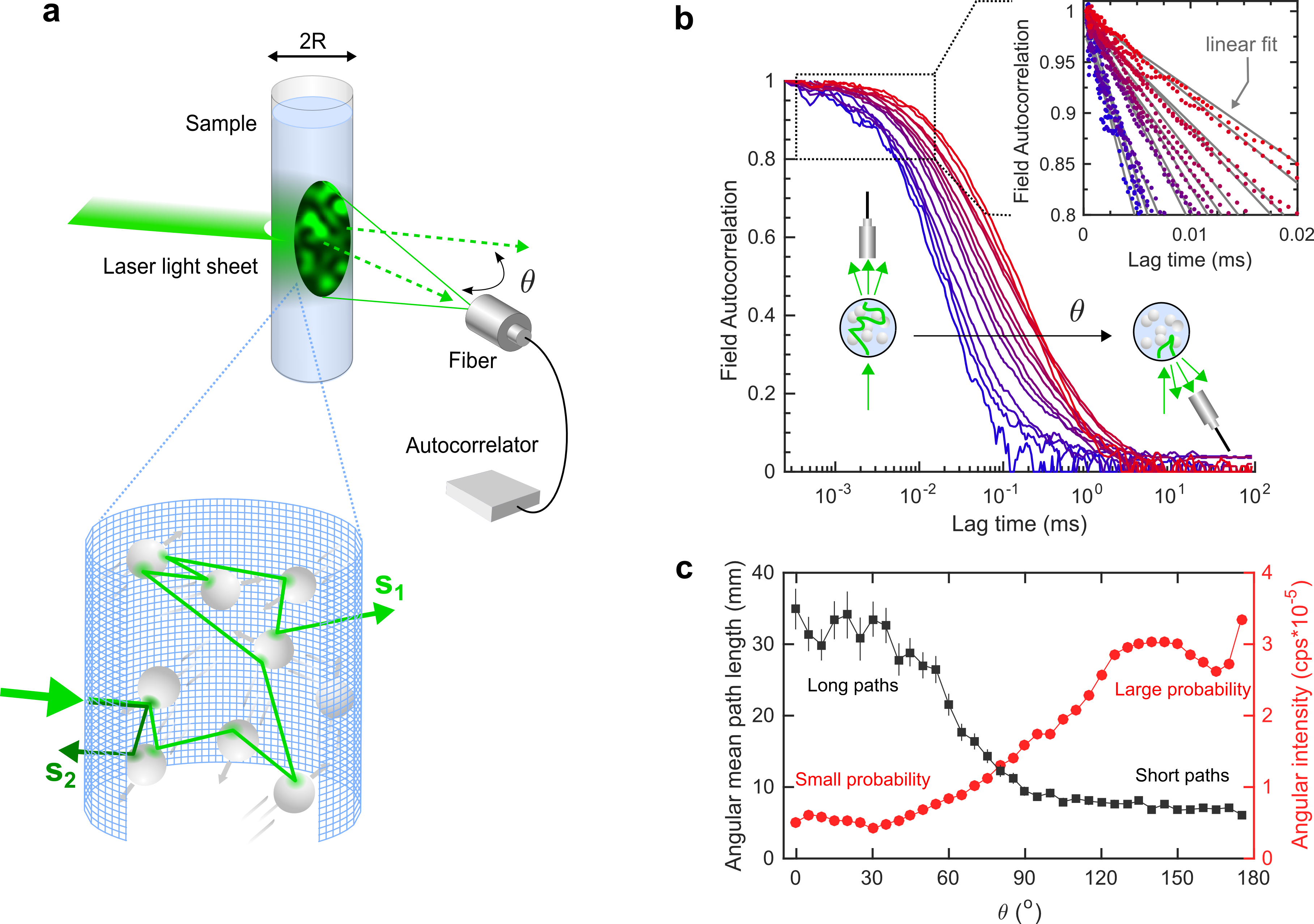}
   \caption{Measurement of the mean optical path length in dynamic disordered media.~\textbf{a}, Illustration of the
   experimental setup. Samples consist of water-dispersed polystyrene micro-beads contained in a cylindrical glass cell of
   internal diameter $2R=\SI{8.54}{mm}$. The cell is isotropically illuminated with a horizontal laser light sheet and the
   speckle pattern of the scattered light is collected with a multimode fiber placed far from the sample. The sample-fiber distance is set to have a
    field of view covering the complete vertical extension of the scattered light. 
 Trajectories are schematically represented as long trajectories leaving the sample in transmission ($\mathrm{s}_1$) and short trajectories leaving the sample in reflection ($\mathrm{s}_2$). To measure the contribution to the mean path length from all trajectories the fiber rotates around the sample to
   sequentially collect light from all output positions and with all radial directions (from $\SI{0}{\degree}$ to
   $\SI{175}{\degree}$). At each angle the collected light is guided to an autocorrelator for the analysis of the temporal intensity
   fluctuations, from which the field autocorrelation is retrieved.~\textbf{b}, Measured electric field autocorrelation at increasing angles 
   ($x$-axis in log-scale). For clarity only some selected curves are shown (at $0$, $40$, $45$, $55$, $60$, $65$, $70$,
   $75$, $80$, $85$, $90$, $95$, $170$, $175$ degrees). The large variation of decorrelation times is caused by the
   strong dependence of the  optical path length on the collection angle. Here, longer paths dominate at small angles (in transmission) and shorter paths at large
   angles (in reflection).  For each angle, the value of the mean optical path length is extracted from the derivative at the origin of the autocorrelation curves, measured here as the slope of a linear fit at very early times (see inset).~\textbf{c}, Angle-dependent mean optical path length (black square) and corresponding scattered light intensity (red circles) measured at steps of $\SI{5}{\degree}$. 
   Error bars give $\SI{95}{\%}$ statistical confidence and are calculated by propagating the experimental errors that
   contribute, i.e. the error on the linear fit performed in \textbf{b} and errors on the measurements of $\ell^*$ and $D$. Errors on
   intensities are of a few per cent and appear smaller than the markers. The global mean path length is obtained by averaging the black squares with the red circles as weighting factors.}
   \label{fig:figure2}
\end{figure*}

\begin{figure*}[t!]
   \centering
   \includegraphics[scale=1]{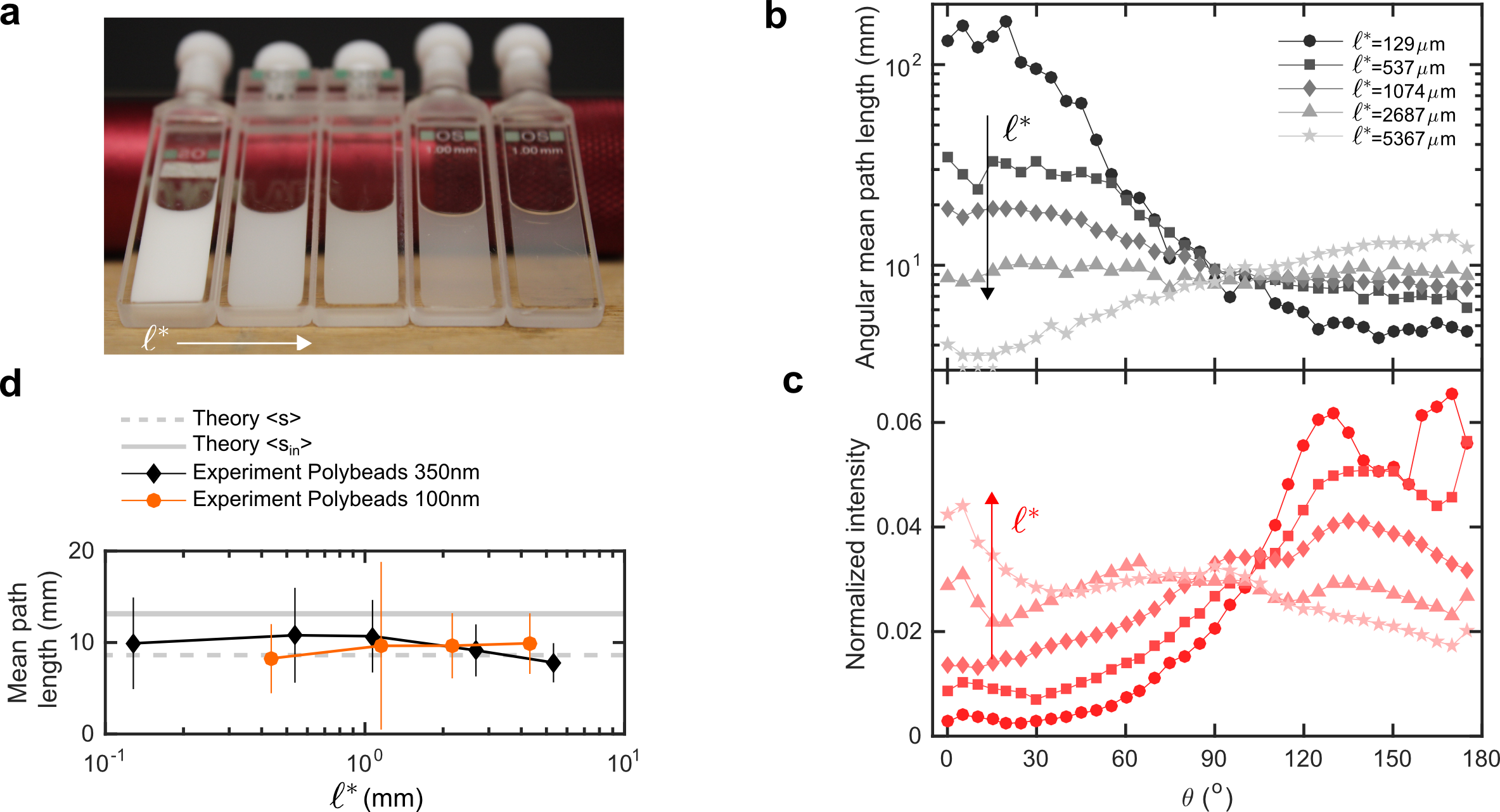}
   \caption{ Observation of the invariance at different scattering strengths.~\textbf{a}, Colloidal solutions with different beads concentration used in the experiment, covering nearly two orders of magnitude in
   scattering strength. The photograph shows $\SI{1}{mm}$ thick slab cuvettes in which these solutions appear transparent or
   opaque, depending on the absolute optical thicknesses in the cylindrical cell ($D/\ell^*$) ranging from about $70$ to
   $1.5$. The arrow indicates increasing transport mean free path, i.e. decreasing scattering strength.~\textbf{b}, Angular distribution of measured mean path length when varying the sample concentration ($y$-axis
   in log-scale). Error bars are not
   displayed given the log-scale, the typical uncertainty is shown in Fig.~\ref{fig:figure2}\textbf{c}.~\textbf{c}, Angular distribution of scattered intensity for
   varying sample concentration. Markers
   refer to the same transport mean free path as in \textbf{b}.~\textbf{d}, Measured mean optical path length as a
   function of the transport mean  free path. Each point represents the mean path length measured over all output
   channels and is obtained as the average over the corresponding angular distribution (\textbf{b}) with the intensity (\textbf{c}) as a weighting
   function. The horizontal dashed gray line indicates the expected value calculated
   based on pure geometrical considerations, $\langle s \rangle=2R$, where $R$ is the radius of the
   cylindrical surface of the samples. Because of the glass cell, the expected value is slightly modified and can be computed by Monte-Carlo simulations (see Supplementary Information) : $\langle s_{\text{in}}\rangle$ is represented by a horizontal solid gray line and corresponds to the average
   path length restricted to the scattering region, including all boundary effects.  Black diamonds represent measurements on polystyrene particles with a diameter of
   $\SI{350}{nm}$ and anisotropy factor $g\approx 0.8$. Orange circles represent measurements on polystyrene particles
   with a diameter of $\SI{100}{nm}$ and anisotropy factor $g\approx 0.1$. Both types of samples show remarkable
   agreement with the numerical prediction for the path length invariance. Error bars give $99\%$ statistical confidence
   and are calculated by propagating the experimental errors that contribute, i.e. the error on the linear fit performed in Fig.~\ref{fig:figure2}b, errors on the measurements of $\ell^*$ and $D$.} 
   \label{fig:figure3}
\end{figure*}

In stark contrast with this view, a recent theoretical study pointed out that a very
fundamental property of wave transport is completely insensitive to the structure of the underlying medium~\cite{Pierrat:2014bp}.
Specifically, it was shown that under very general assumptions the mean path length associated with wave scattering
through a medium only depends on the medium's boundary geometry, but not on its internal microstructure. To arrive at
this result, an invariance property first found for random walks~\cite{Blanco:2003tm} was generalized to arbitrary wave
scattering scenarios based on early insights from the pioneers of 20th century physics like Weyl, Wigner, Krein and
Schwinger. As such, this invariance relation is completely general and holds for the movement of ants through a
designated two-dimensional area just as well as for the propagation of light waves through a disordered material (see
Fig.~\ref{fig:figure1}). In all cases the mean length $\langle s\rangle$ of trajectories that enter the medium
at arbitrary positions and incident angles up to the point where they exit the medium is predicted to be independent of
whether the medium is uniform, highly structured, disordered or anything in between. It is found to be
\begin{equation}\label{eq:path_length}
   \langle s\rangle=v_E \langle t\rangle=\frac{4V}{\Sigma}
\end{equation}
for a three-dimensional geometry of volume $V$ and surface $\Sigma$~\cite{Pierrat:2014bp}. Here, $v_E$ is the transport
velocity, which for waves transport in resonant media takes into account the dwell time inside the particles (see Supplementary Information).
For the paradigmatic case of a fully disordered medium the crossover between systems with different degrees of disorder
can be conveniently described by the transport mean free path  $\ell^*$, that corresponds to the length after which the
propagation direction of an incoming wave or particle gets randomized. Applying the theoretical predictions to this case
would mean that a change of $\ell^*$ should leave the mean path length invariant. To experimentally demonstrate this surprising theoretical result, we investigate multiple scattering of light in a colloidal suspension of particles in water (see Fig.~\ref{fig:figure2},~\ref{fig:figure3}). By varying the
concentration and size of the particles, we tune the mean free path by almost two orders of magnitude, covering the
range of a nearly transparent to a very opaque system.  We measure the mean length of light trajectories from  temporal decorrelation of an optical speckle pattern in each one of these suspensions, and unambiguously observe this invariance. Quite remarkably, the
\emph{distribution} of path lengths gets modified significantly when changing the transport mean free path - only the
\emph{mean value} of the distribution stays unchanged.\\
\textit{Experiment.}
When shining light on a disordered medium the spatial inhomogeneity of the refractive index prohibit a straight-line propagation,
forcing the wave instead to scatter in all available directions, see Fig.~\ref{fig:figure1}c. To measure the resulting
optical path length distribution $P(s)$, the most common method uses ultrashort pulses and time-resolved detection
schemes~\cite{kop1997observation, pattelli2016spatio}. However, we do not require access to the full distribution $P(s)$, but just to its mean value $\langle s\rangle =
\int_0^\infty P(s)s ds$. We therefore developed a novel technique derived from DWS (Diffusing Wave
Spectroscopy)~\cite{maret1987multiple, pine1988diffusing} to directly measure the mean optical path
length in dynamic scattering media with high sensitivity and dynamic range. In this approach we  illuminate the sample with a monochromatic laser and measure the autocorrelation function $g_E(\tau)= | \langle E^*(\tau) E(0)
\rangle|^2/| \langle E E^*\rangle |^2$ of the temporal fluctuations in the scattered light field $E(t)$. We exploit
the intimate connection between speckle fluctuations and the distribution of optical path lengths $P(s)$, which is
formalized as
\begin{equation}
   \label{eq:field_autocorr}
   g_E(\tau)=\int_0^\infty ds P(s) \exp\left(- \frac{s}{\ell^*}\frac{2\tau}{\tau_0}\right),
\end{equation}
where $\tau_0=1/(k^2D)$, $k$ is the wave vector inside the medium and $D$ is the diffusion constant of the scattering particles. Quantitative information on the mean value can be immediately retrieved by considering the very early-time
decorrelation. 
Indeed, the derivation of Eq.~\ref{eq:field_autocorr} evaluated at \(\tau=0\) leads to
an explicit expression for the mean optical path length
\begin{equation}
   \label{eq:meanpathlength}
   \langle s \rangle=-\left.\frac{dg_E}{d\tau}\right\vert _{\tau=0}  \frac{\ell^* \lambda_0^2}{8\pi^2n^2D},
\end{equation}
where $\lambda_0$ is the light wavelength in vacuum and $n$ is the medium refractive index.  Note that  Eqs.~(\ref{eq:field_autocorr}) and~(\ref{eq:meanpathlength}) rely  only on the so-called continuum approximation of the multiple scattering process \cite{durian1995accuracy},  which in practice requires a few  scattering events to be valid~\cite{durian1995accuracy, BOAS-1998,CARMINATI-2004}. Here, since even paths with very few scattering events can contribute to the mean path length, the accuracy
of Eqs.~(\ref{eq:field_autocorr}) and~(\ref{eq:meanpathlength}) has been additionally verified for all experimental situations by using Monte Carlo simulations (see
Supplementary Information).

The experiment is illustrated in Fig.~\ref{fig:figure2}a.  The scattering solution is contained in a  cylinder glass cell, which supports the liquid and defines its geometrical features. In this elongated geometry the $ V/\Sigma$ ratio simplifies into $R/2$, where $R$ is the cylinder radius, and the expected mean path length from Eq.(~\ref{eq:path_length}) is $\left\langle s
\right\rangle = 2R$.
 Because of the index mismatch between the scattering medium, the glass cell, and the outer
regions, Eq.(~\ref{eq:path_length}) cannot be directly applied because of multiple boundary reflections. We therefore developed a more refined model taking into account correct boundary conditions, which shows that the invariance property remains valid in the presence of interfaces. For the simple case of a single boundary it  leads to $\langle s_{\text{theo}}\rangle=v_E \langle t\rangle=(4V/\Sigma)(n_2^2/n_1^2),$ where $n_1$ ($n_2$) is the refractive index of the outer (scattering) region respectively (see Supplementary
Information).  Here we are in the more complex situation of multiple boundaries (medium-glass-air), furthermore our technique only gives access to the light path inside the scattering region (not in the glass part). Nonetheless,  Monte-Carlo simulations allow us  to find the expected invariant mean path length $\langle s_{in}\rangle$ for this experimental situation. Qualitatively, the presence of boundaries means that light exiting the scattering region has a  probability to be reflected back to propagate again into the scattering medium, therefore $\langle s_{in}\rangle$ is larger compared to the ideal case, where all trajectories hitting the boundaries leave the region. ~\cite{Pierrat:2014bp,Blanco:2003tm}.

We illuminate the sample with a laser light sheet all along its diameter such that, as required to observe the
invariance,
light enters  with all possible angles with respect to the surface normal. To collect light from all surface locations, we use a multimode fiber ($\SI{10}{\micro\meter}$ core) mounted sufficiently far away from the sample, and placed on a goniometric
mechanical arm for angle-resolved measurements. The multimode fiber guides light to single photon counters, and a coincidence electronics allows us to measure the temporal autocorrelation. 
In Fig.~\ref{fig:figure2}b we show an example of measurements for a sample with $\ell^* \approx \SI{500}{\micro\meter}$  for detection
angles ranging  from $\SI{0}{\degree}$ (forward direction) to $\SI{175}{\degree}$ (backward direction). For this rather opaque sample, $2R/\ell^* \approx 20$, the  characteristic decorrelation time of the autocorrelation function ranges from tens of
\si{\micro\second}  in the forward direction to some \si{\milli\second} in the backward direction. Since the decorrelation time decreases as the number of scattering events increases, our results provide evidence of the significant variation in the average number of scattering events between transmitted and reflected  light. We adopt Eq.~(\ref{eq:meanpathlength}) to measure the mean optical path length at each angle and evaluate the derivative  of the autocorrelation at the origin as the
slope of  a linear fit of the first measured points, as shown in the closeup of Fig.~\ref{fig:figure2}b. The multiplicative coefficient
in Eq.~(\ref{eq:meanpathlength}) containing  $\ell^*$ and $D$ is measured  with an independent characterization of the
sample (see Supplementary Information). The recorded angular mean path lengths are shown in
Fig.~\ref{fig:figure2}c, together with the corresponding  scattered light intensity,  which quantifies the probability
to have light exiting in this particular direction. These measurements show that long  trajectories, which contribute to the overall mean with large values,
are less probable than short trajectories, which in turn are more abundant, but contribute to the overall mean with small values. This
feature illustrates the delicate balance between long and short trajectories that enables the mean value to be
independent of the actual path length distribution and which is at the root of this invariance property.\\ 

\textit{Results.}
The most striking feature appears when considering the variation of both the angular mean path length and the corresponding
intensity for different values of the transport mean free path $\ell^*$, shown in Fig.~\ref{fig:figure3}b and
Fig.~\ref{fig:figure3}c. Starting from the most opaque sample,  and decreasing  the scattering strength (i.e., for increasing transport mean free path), both the
path length curve as well as the associated intensity get more symmetrically distributed among all angles. Then, when
the transport mean free path becomes comparable to the sample size, the symmetry of the angular mean path length curve becomes
completely inverted as the sample becomes more transparent and longer path lengths are observed in reflection rather than in transmission. The symmetry of the
intensity in turn is inverted in the opposite sense such as to completely compensate the redistribution of path lengths.
In our analysis, we measure the overall mean path length over all trajectories, therefore these two distributions get convolved in a weighted angular average. This average is thus  evaluated by multiplying the angle-dependent values of Fig.~\ref{fig:figure3}b with the intensities of
Fig.~\ref{fig:figure3}c as weighting functions: $\langle s_{exp} \rangle=\sum_i \langle s \rangle(\theta_i) I(\theta_i)/\sum_i
I(\theta_i)$, where the index $i$ indicates the angle of measurement. The striking result  we obtain, which is the main result of the paper,  is shown in Fig.~\ref{fig:figure3}d: The measured mean path length stays invariant over nearly two orders of magnitude of transport mean free path. This is in remarkable agreement with
the numerical prediction taking into account the real geometry of the system (i.e. glass cell with two interfaces). We observe a small deviation for very weakly scattering sample, where we expect the invariance to hold, but where the model underlying our measurement starts to fail. Furthermore, the
particles we used have a pronounced scattering anisotropy. To test this invariance property also on optically very different systems, we repeated the experiment
using smaller colloidal particles with diameters of about $\SI{100}{nm}$ corresponding to almost isotropic scattering (see orange curve in Fig.~\ref{fig:figure3}d). Also here the
invariance is verified, confirming that neither the transport mean path nor the anisotropy affect this robust property.\\

\textit{Discussion and conclusions.}
In summary, we provide the first experimental demonstration of a novel and universal invariance property for wave
scattering in disordered media. Since the path lengths in a medium are intimately connected with a variety of other
crucial concepts, like the absorption~\cite{Muskens:2008kn,Garnett:2010en,vynck2012photonmenagement}, the dwell time~\cite{LAGENDIJK-1996} and the frequency robustness of states in this medium~\cite{carpenter2015observation,xiong2016spatiotemporal}, we expect
the invariance property established here to set very rigid bounds on what can be achieved by modifying the underlying
medium structure. 
Implications are particularly obvious for light harvesting, light deposition and imaging techniques~\cite{polman2012photonic,boriskina2016roadmap,ntziachristos2010going}.
We also emphasize that the path length invariance is neither restricted to light propagation nor to random walks, but
applies basically to all wave scattering problems, ranging from matter waves on the smallest length scales to
gravitational waves on the largest conceivable dimensions. As such our demonstration provides just a first glimpse onto
the many different contexts in which this type of physics plays a role.
\section{Acknowledgement}
Authors wish to thank J. Schwarz, P. Ambichl, A. Haber and J. Bertolotti for fruitful discussions and T. Narita, F. Pincet and C. Francois-Martin for technical assistance with the DLS machine. This work was supported by the European Research Council (Proj. Ref. 278025). R.P. and R.C. were supported by LABEX WIFI (Laboratory of Excellence within the French Program ``Investments for the Future'') under references ANR-10-LABX-24 and ANR-10-IDEX-0001-02 PSL*. S.R. was supported by the Austrian Science Fund (FWF) through Project No. SFB NextLite F49-P10.
\section{Author contributions}
R.S. and S.G. conceived the experiment. R.S. and U.N. carried out the experiment. R.S. and R.P. performed simulations. R.P., R.C. and S.R. devised the theoretical frame work and carried out the corresponding calculations. All authors discussed results and contributed to the writing of the paper.
\section*{Correspondence}
Correspondence should be addressed to R.S.~(email: romolo.savo@lkb.ens.fr) and/or to S.G.~(email: sylvain.gigan@lkb.ens.fr).

\bibliographystyle{unsrt}
\bibliography{bibliography.bib}

\end{document}


\title{SUPPLEMENTAL MATERIAL: Mean path length invariance in multiple light scattering}

\author{Romolo Savo}
 \affiliation{%
Laboratoire Kastler Brossel, UMR 8552, CNRS, Ecole Normale Sup\'{e}rieure, Universit\'{e} Pierre et Marie Curie, Coll\`{e}ge de France, 24 rue Lhomond, 75005 Paris, France
}%
\author{Romain Pierrat}
\affiliation{
 ESPCI Paris, PSL Research University, CNRS, Institut Langevin, 1 rue Jussieu, 75005, Paris, France
}%
\author{Ulysse Najar}
\affiliation{%
Laboratoire Kastler Brossel, UMR 8552, CNRS, Ecole Normale Sup\'{e}rieure, Universit\'{e} Pierre et Marie Curie, Coll\`{e}ge de France, 24 rue Lhomond, 75005 Paris, France
}%
\author{R\'emi Carminati}
\affiliation{
 ESPCI Paris, PSL Research University, CNRS, Institut Langevin, 1 rue Jussieu, 75005, Paris, France
}%
\author{Stefan Rotter}
\affiliation{
 Institute for Theoretical Physics, Vienna University of Technology (TU Wien), Vienna, A-1040, Austria
}%
\author{Sylvain Gigan}
 \affiliation{%
Laboratoire Kastler Brossel, UMR 8552, CNRS, Ecole Normale Sup\'{e}rieure, Universit\'{e} Pierre et Marie Curie, Coll\`{e}ge de France, 24 rue Lhomond, 75005 Paris, France
}%

\maketitle
\section*{Experimental setup}
\begin{figure}
   \centering
   \includegraphics[scale=1]{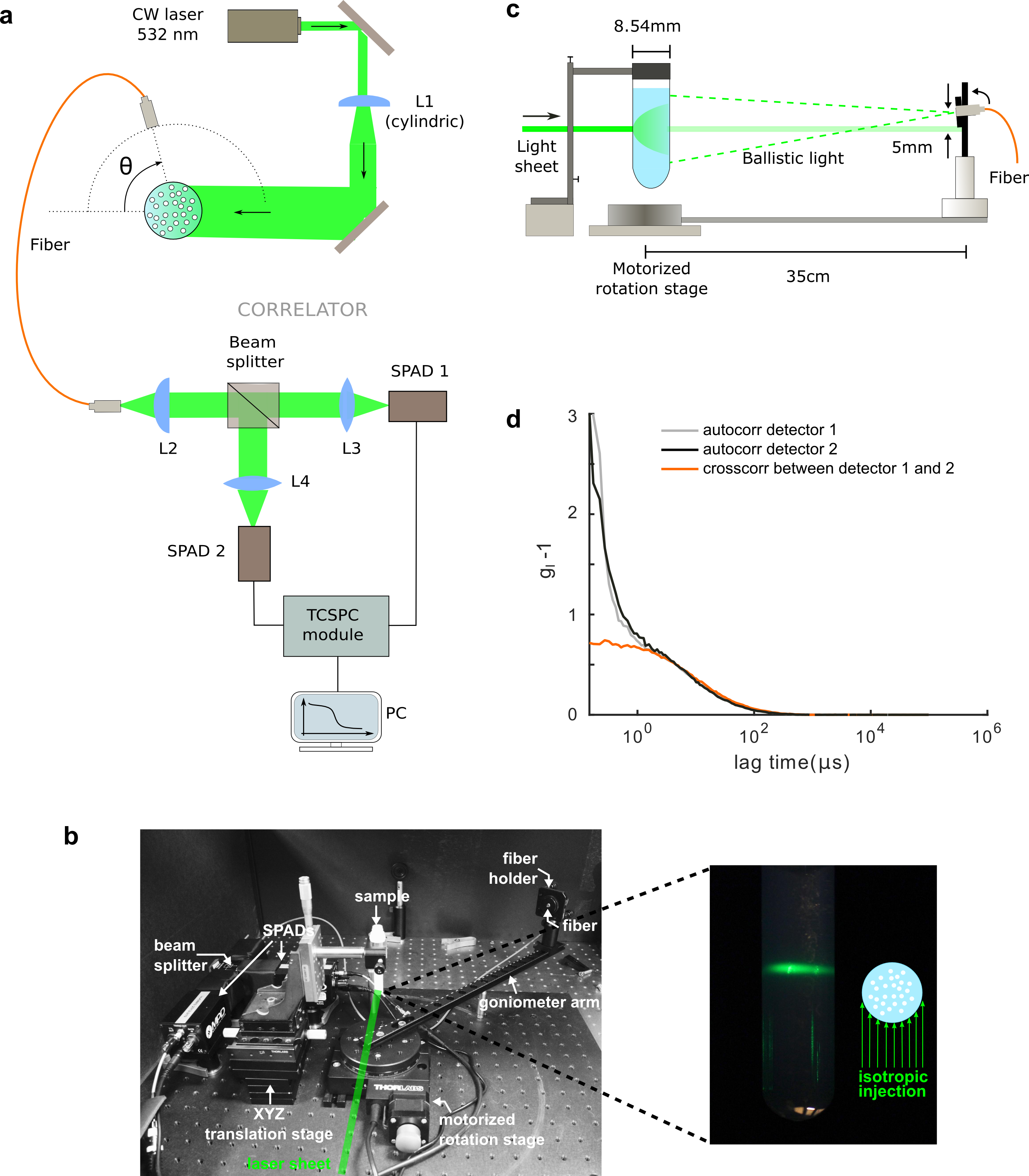}
   \caption{Experimental setup. \textbf{a}, Schematic illustration of the experimental setup. \textbf{b}, Table-top view of the experiment. The closeup highlights the isotropic light injection in the sample via a light sheet illumination on its cylindrical surface. \textbf{c}, Detailed configuration of light collection geometry. \textbf{d}, Example of the effect determined by after-pulsing at early times  on the intensity autocorrelation measured with a single-detector (black and gray lines). This effect is removed by measuring the cross-correlation between the two detectors (orange line).}
   \label{fig:S_setup_1}
\end{figure}

A schematic of the experimental setup is shown  in Fig.~\ref{fig:S_setup_1}a together with a picture of  the table-top implementation in Fig.~\ref{fig:S_setup_1}b\footnote{All references in this text refer to the Supplemental Material if not differently specified}.
Illumination is provided by a (CW) laser (COHERENT Sapphire) at $\lambda=\SI{532}{nm}$, $\SI{50}{mW}$, with vertical polarization. The generated narrow  beam is horizontally stretched (on a plane parallel to the table) by means of a cylindrical lens ($f=\SI{1}{m}$). The beam  arrives on the sample strongly asymmetrically (light sheet), with an horizontal width of $\SI{8}{mm}$ and thickness less than $\SI{1}{mm}$.  Given the circular section of  the sample such illumination provides  a uniform and isotropic light injection on the horizontal plane, see closeup in Fig.~\ref{fig:S_setup_1}b.  

Light is collected through a multimode optical fiber (diameter \SI{10}{\micro\meter}, $NA=0.1$, THORLABS), mounted on a computer controlled rotating arm to radially  scan  scattered light around the sample.  The detection configuration is detailed in Fig.~\ref{fig:S_setup_1}c. The fiber is placed far from the sample with a field-of-view that completely covers the diffuse halo on the sample surface.  This  guarantees collection from  all light trajectories at a fixed angle.  Given the wide range of sample turbidities, from  completely opaque to almost transparent, the presence of ballistic light and of the stray light created by reflections is a crucial aspect to consider.  In particular in diluted samples and  at small angles, e.g. $\SI{0}{\degree}$ to $\SI{5}{\degree}$, ballistic  and stray light  may completely cover the  scattered signal.  Moreover, even if  field fluctuations are still observable, their interference with a ballistic (i.e. stationary)  field  alters the signal autocorrelation function \cite{cipelletti1999ultralow}. To avoid collection of ballistic light, the fiber tip is placed above the beam propagation plane and then tilted to keep its field-of-view  always matched with the  sample diffuse halo.

Collected light  is guided through the fiber to  the  optical correlation part, see Fig.~\ref{fig:S_setup_1}a. Lenses $\mathrm{L}2$, $\mathrm{L}3$ and $\mathrm{L}4$ have focal lengths respectively of $\SI{50}{mm}$, $\SI{25}{mm}$ and $\SI{25}{mm}$.  The beam splitter has a $50:50$ ratio. Detectors  are Single Photon Avalanche Diodes (SPAD) (MPD-Micro Photon Devices) working in the photon counting regime. 
We employ a double-detector  configuration  to measure  the cross-correlation of the two identical signals instead of the  single signal  auto-correlation. This configuration allows avoiding after-pulsing  artifacts typical of SPADs and obtaining sub-$\mu$s temporal resolution~\cite{hobel1994dead}, see Fig.~\ref{fig:S_setup_1}d.
Signals from the SPADs are sent to two independent  channels of a  Time-Correlated Single Photon Counting (TCSPC) module (PicoQuant HydraHarp400) for measuring their cross-correlation \textit{on-the-fly}.

\section*{Connection between the field autocorrelation and optical path lengths}
\begin{figure}[t]
   \centering
   \includegraphics[scale=0.95]{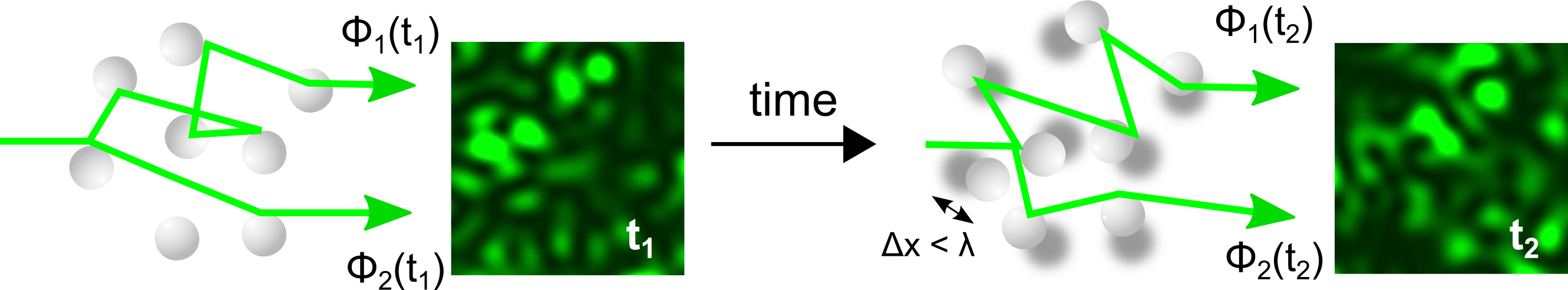}
   \caption{Illustration of the phase accumulation mechanism along two generic light paths in a dynamic disordered medium at different instants. All scattered trajectories interfere locally and instantaneously in the medium generating a speckle pattern, which is characterized by a complex distribution of bright and dark spots. Shown images of speckles are illustrative. Thermal motion of the scatterers forces trajectories to be redefined and make the speckle decorrelate in time. In the deep multiple scattering regime the speckle fully decorrelates for particle displacements that are smaller than the wavelength. }
   \label{fig:S_DWS_trajectories_decorr}
\end{figure}
Elastic scattering between (coherent) laser light and micro-particles in a
colloidal solution creates the complex interference pattern known as \textit{speckle}, see images in Fig.~\ref{fig:S_DWS_trajectories_decorr}. Bright and dark spots in the speckle represent, respectively, the constructive and destructive
interference of several light fields that have traveled along different trajectories in the medium and have
accumulated a different phase. When the scattering centers move, e.g. through thermal fluctuations, light trajectories are
instantaneously redefined and  the locations of the constructive and destructive interference spots change accordingly.
The net result is a time-dependent speckle pattern, whose temporal features depends on the distribution of optical path
length in the medium and on the particles' diffusion constant~\cite{maret1987multiple,pine1988diffusing}. More precisely, the temporal field autocorrelation in a speckle spot is the Laplace transform of the distribution of all optical path length that are interfering in the spot. This is the property expressed by Eq.~(2) in the main text. 

\section*{Field autocorrelation from TCSPC intensity measurements}
\begin{figure}[!t]
   \centering
   \includegraphics[scale=1]{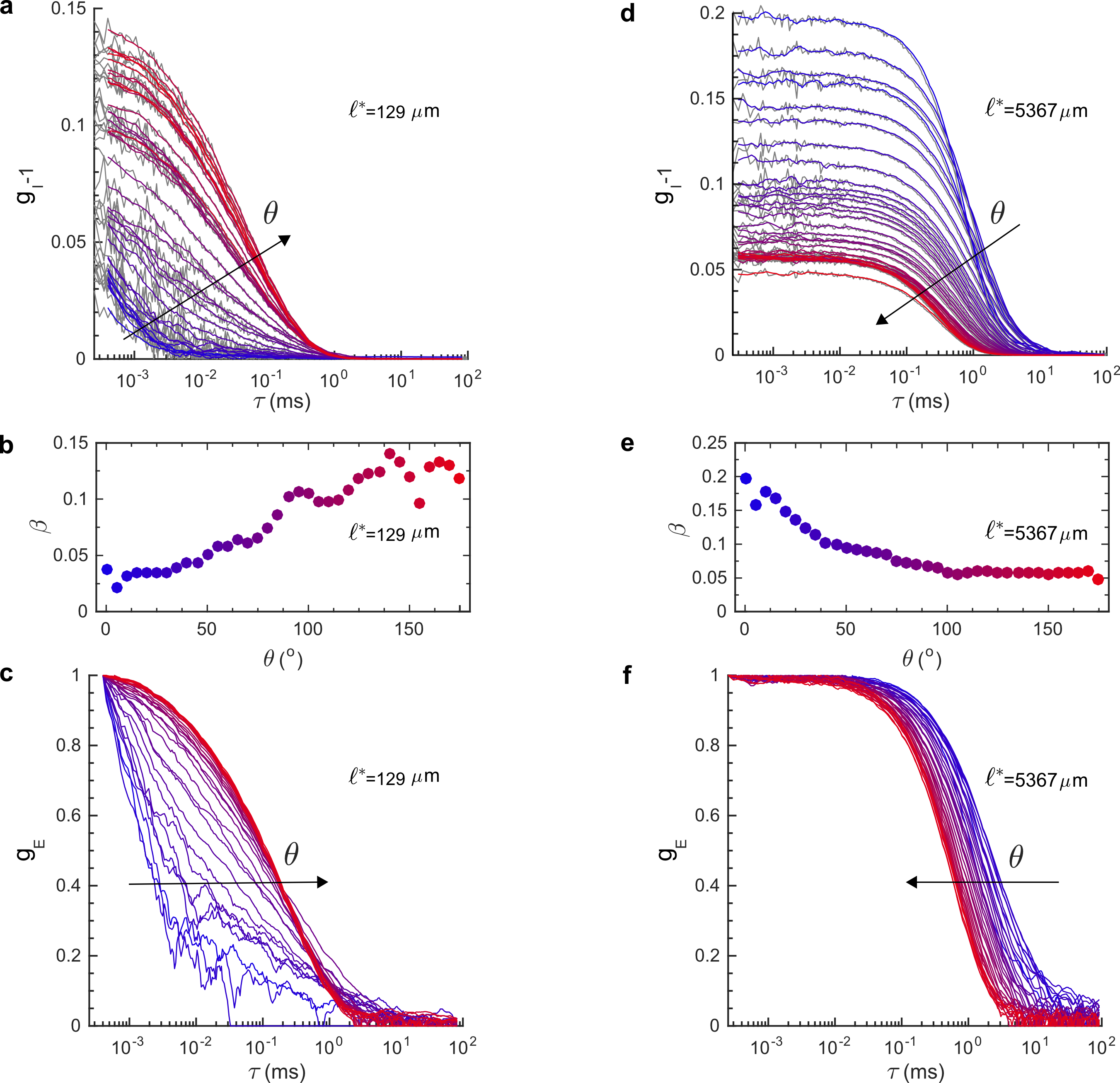}
   \caption{Retrieval of the field autocorrelation $g_E$ from raw measurements of the intensity autocorrelation $g_I$. \textbf{a}-\textbf{b}-\textbf{c} refer to measurements on the most strongly scattering sample (polybeads 350nm, $\ell^*=\SI{129}{\micro\meter}$). \textbf{d}-\textbf{e}-\textbf{f} refer to measurements on the least strongly scattering sample (polybeads 350nm, $\ell^*=\SI{5367}{\micro\meter}$). Color scale (from blue to red) indicates increasing collection angles, from transmission to reflection.}
   \label{fig:S_GI_to_GE}
\end{figure}
The TCSPC module measures the arrival time $t_i=t_0, t_1, t_2,..., t_n $ of each photon over the entire measurement time. A histogram of this time-sequence provides a  measurement $I(t)$ of the speckle intensity fluctuation, where the time resolution is set by the binning interval $\Delta t$. By definition, the intensity autocorrelation function is given by $g_I(\tau) = \langle I(\tau) I(0) \rangle/\langle I(0)\rangle^2$, where $\tau$ is the lag time. However, our TCSPC module directly provides an \textit{on-the-fly} calculation of $g_I(\tau)$  based on the hardware implementation of a multi-$\tau$ algorithm~\cite{wahl2003fast}, which allows us to span several decades on the temporal axis (from hundreds of $\SI{}{ns}$ to hundreds of $\SI{}{ms}$).

We retrieve the autocorrelation of the speckle field from the measured intensity autocorrelation via the Siegert relation ~\cite{berne2000dynamic,lemieux1999investigating} 
\begin{equation}
g_E(\tau)= \sqrt{ \frac{g_I-1}{\beta}},
\label{eq:S_Siegart_relation}
\end{equation}
where $0\leq\beta\leq1$ is a geometric factor determined by the collection optics, which quantifies the
number of speckle spots simultaneously collected by the fiber tip. Note that collecting multiple speckle spots does not vitiate the autocorrelation  measurement, but it only  reduces the optical signal-to noise ratio~\cite{durduran2014diffuse}.
As $g_E(0)=1$ by definition, the parameter $\beta$ can be determined as $\beta=g_I(0)-1$~\cite{cipelletti1999ultralow}. The obtained value is then used in Eq.~(\ref{eq:S_Siegart_relation}) to retrieve the normalized field autocorrelation function. 

In Fig.~\ref{fig:S_GI_to_GE} we show essential steps to retrieve $g_E$ from the measured $g_I$. To highlight the optimal performance of the setup, e.g. sensitivity and dynamic range, we show two limiting cases from our set of measurements, namely for the most strongly scattering ($\ell^*=\SI{129}{\micro\meter}$) and least strongly scattering ($\ell^*=\SI{5367}{\micro\meter}$) sample. 
In Fig.~\ref{fig:S_GI_to_GE}a and Fig.~\ref{fig:S_GI_to_GE}d the solid gray line corresponds to raw measurements of $g_I$. These measurements are convoluted  with a narrow rectangular function to smooth noise oscillations and obtain the solid colored curves. In Fig.~\ref{fig:S_GI_to_GE}b and Fig.~\ref{fig:S_GI_to_GE}e we report the angular dependence of the corresponding $\beta$ values. For most strongly scattering samples, Fig.~\ref{fig:S_GI_to_GE}b, we observe that $\beta$ increases with the angle, while this trend is inverted for the least strongly scattering sample, Fig.~\ref{fig:S_GI_to_GE}e. This can be understood by considering that the size of the single speckle spot on the fiber tip depends uniquely on the vertical extension of the diffuse halo, as the sample-fiber distance is fixed~\cite{goodman1976some}. More precisely measurements show that the extension of the diffuse halo in our cylindrical geometry decreases with the angle for opaque samples (strong scattering) and increases with the angle for almost transparent samples (weak scattering). In Fig.~\ref{fig:S_GI_to_GE}c and Fig.~\ref{fig:S_GI_to_GE}f we show the retrieved field autocorrelation. Even if these measurements correspond to two limiting cases for our instrumentation, i.e. very fast decorrelation and very slow decorrelation,  all field autocorrelations are well normalized, continuous at short lag times and correctly drop to zero at long times.      

\section*{Early-time fit of the field autocorrelation }
\begin{figure}[t]
   \centering
   \includegraphics[scale=1]{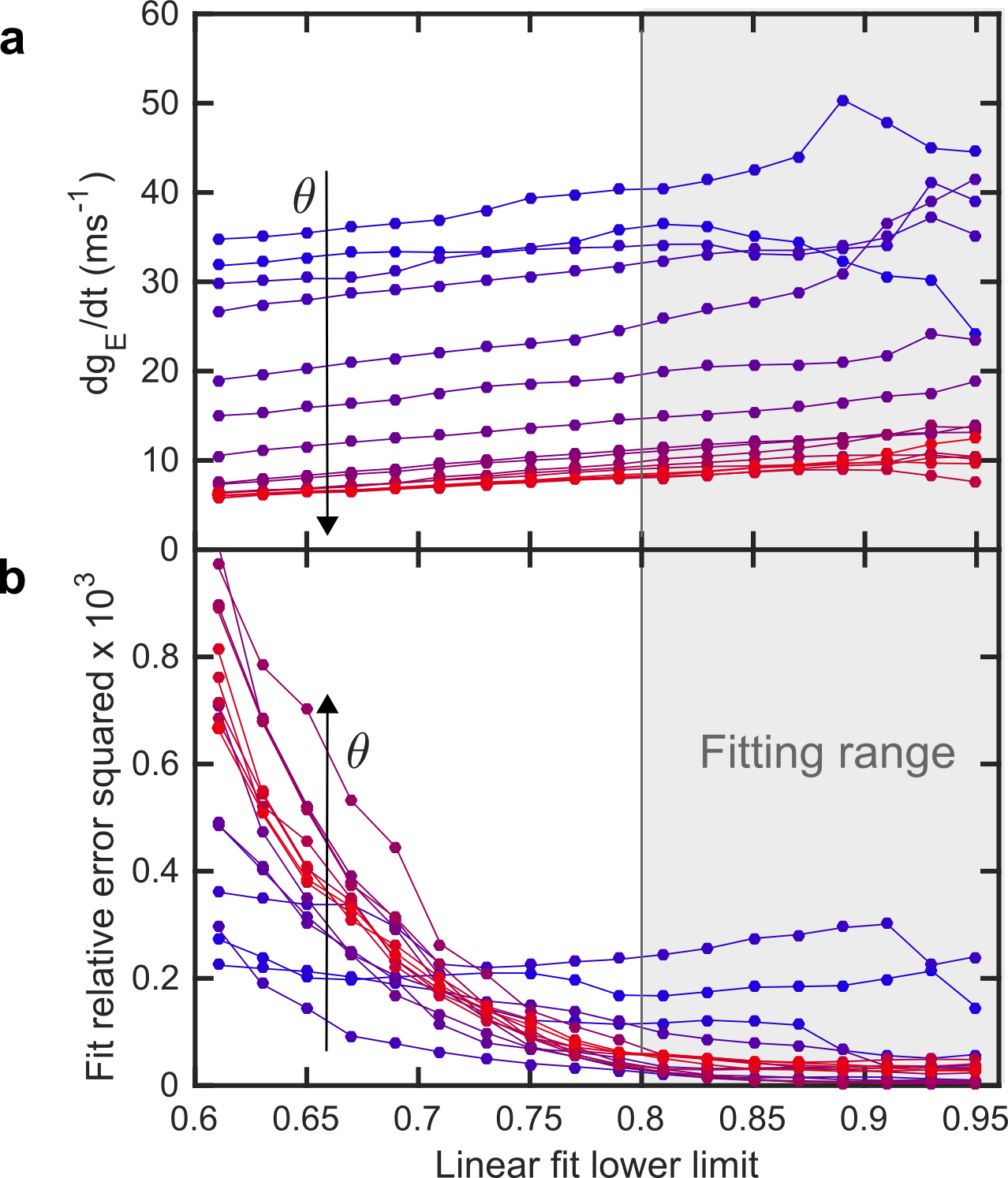}
   \caption{Strategy for the definition of the interval of linear fit at early times. \textbf{a}, Slope of the linear fit when varying the lower limit of the fitting interval. The upper limit is fixed to $g_e=1$ \textbf{b}, Relative error squared of the linear fit in the same circumstances. Color scale (from blue to red) indicates increasing angles. The gray region indicates the adopted fitting interval.}
   \label{fig:S_Linear_fit_lower_limit}
\end{figure}
Early-time features of the field autocorrelation contain information on the mean length of collected light paths. As expressed by Eq.~(3) in the main text, the mean path length is proportional to $dg_E/d\tau$ at $\tau=0$. To measure this quantity we perform a linear fit on the \textit{early times } of $g_E$ and take its slope as the derivative. Such a first-order approximation is very robust since, even in the multiple scattering regime, the autocorrelation is expected to decay quasi-exponentially at the beginning~\cite{maret1987multiple,pine1988diffusing,lemieux1999investigating}. Moreover, by calculating the slope over a set of points~\textendash~and not only over the first two or three points~\textendash~we obtain  a value that is minimally sensitive to noise fluctuations. 

As the autocorrelation decay is heavily  dependent on the scattering mean free path and on the detection angle, see Fig.~\ref{fig:S_GI_to_GE}, typical \textit{early times } may vary significantly. This feature makes it impossible to define a common time range for the linear fit. To find a general rule for the definition of \textit{early times} we consider the absolute value of the field autocorrelation instead of the lag times. Indeed, we see that the time interval corresponding to the decay in which $1\leq g_E \leq 0.8$ allows  for very good linear fits regardless of the sample and the collection angle.  
In Fig.~\ref{fig:S_Linear_fit_lower_limit} we consider the sample with $\ell^*=$\SI{500}{\micro\meter} as an example. We report the measured slope as well as the relative error squared of the fit for several choices of the fitting interval. We fix the fitting upper bound at $g_E=1$ and vary the lower bound. For each value of the lower bound the relative error squared is calculated as $\sum_i ((Y_i-O_i)^2/Y_i^2)/N$, where $Y_i$ is the theoretical value of the linear model, $O_i$ is the observed point and $N$ is the number of points in the fitting interval.

In Fig.~\ref{fig:S_Linear_fit_lower_limit}a we see that measured slopes are stable as long as the lower fitting bound is larger than 0.8, while they deviate from the initial value for lower bounds below 0.8. Evidently the autocorrelation decay starts to deviate from the linear behavior at this point. This is confirmed by the trend of the relative error squared in Fig.~\ref{fig:S_Linear_fit_lower_limit}b, which is minimal for a lower bound above 0.8 while quickly grows for fits that includes points below $g_E=0.8$. 

\section*{Samples characterization}
Sample's solutions are prepared by controlled dilution of monodispersed aqueous polystyrene colloids (polybead\textregistered microspheres, Polysciences Inc.) with nominal particle size of \SI{350}{nm} and \SI{100}{nm}. A further characterization of particle's sizes has been performed through the measurement of the particles' diffusion constant. Particles' concentration in original undiluted solutions is respectively $2.6\%$ and $2.7\%$ w$/$v. 
\subsection*{Transport mean free path}
\begin{figure}[t]
   \centering
   \includegraphics[scale=1]{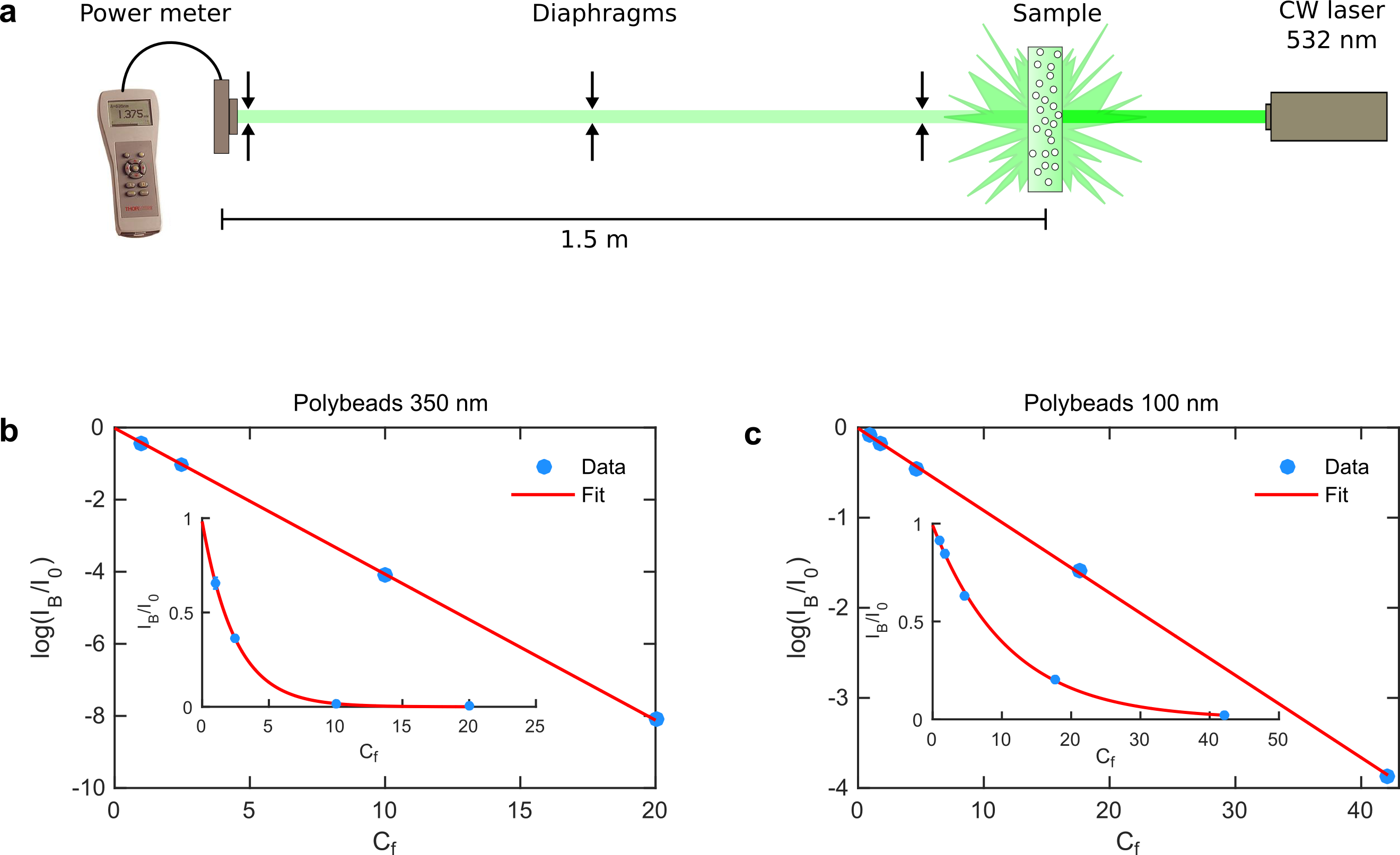}
   \caption{Measurements of the scattering mean free path by means of ballistic transmission measurements. \textbf{a}, Schematic of the experimental setup. \textbf{b}-\textbf{c}, Measured ballistic transmission and relative fits for the two sets of samples. Main figures report ballistic transmissions in logarithmic scale, while insets are in linear scale. Data in the insets contains error bars, which appear always smaller than the markers. Errors on ballistic transmissions is quantified as half of the variability interval for 30 different measurements performed on several positions of the cuvette face.}
   \label{fig:S_ls_measurement}
\end{figure}
In multiple light scattering two important lengths characterize the transport process: one is the scattering mean free path $\ell_s$,  which is the typical free propagation distance between two scattering events; the second is the transport mean free path $\ell^*$, which is the characteristic distance after which scattered light loses memory of its initial direction. In general $\ell^* \neq \ell_s $ when the single scattering process is anisotropic and leaves spare directional correlations between incoming and outgoing light.
We measure the transport mean free path $\ell^*$ of all our samples on the basis of the so called \textit{similarity relation } for light transport~\cite{boas2011handbook,svensson2013exploiting}
\begin{equation}
\ell^*=\frac{\ell_s}{1-g},
\label{eq:similarity_relation}
\end{equation}
which connects the two lengths with the anisotropy factor $g=\langle \cos\theta \rangle$, where $\theta $ is the scattering angle. Normally $0 \leq g \leq 1$, when $g=1$ we have completely forward scattering, when $g=0$ we have completely isotropic scattering and $\ell^* = \ell_s$. We measure $\ell^*$ for all samples and calculate $g$ from Mie Theory. Hence, we use Eq.~(\ref{eq:similarity_relation}) to obtain $\ell^*$.\\ \\
\textit{Scattering mean free path}. We measure the scattering mean free path $\ell_\mathrm{s}$ based on the fact that we have full control of the scatterers'
concentration $C$. We perform accurate weight measurements of the initial colloidal solution and of the added deionized water. By knowing the concentration of the original solution, the particles' size, water and polystyrene density, we retrieve the number of colloidal particles $N$ and the volume of the sample $V$. We obtain the concentration as $C=N/V$ with an uncertainty of about $\SI{0.1}{\%}$. Within each samples set, from the most transparent to the most opaque, we define the concentration factor $\mathrm{C_f}$ as the ratio between the scatterers' concentration of a sample and the scatterers' concentration of the most diluted (transparent) sample. In terms of  $\mathrm{C_f}$ the scattering mean free path $\ell_s$ scales within the sample set as
\begin{equation}
	\ell_s(\mathrm{C_f})=\frac{\ell'_s}{\mathrm{C_f}},
	\label{eq:ls_scaling}
\end{equation}
with $\ell'_s$ being the scattering mean free path of the most diluted sample.\\ 
By using Eq.~(\ref{eq:ls_scaling}) the Beer-Lambert law for the ballistic transmission $I_B$ writes as 
\begin{equation}
   \frac{I_B}{I_0}=\exp\left(-\frac{L}{\ell'_s}\mathrm{C_f}\right),
   \label{eq:BeerLambert_law}
\end{equation}
where $I_0$ is the reference transmission and $L$ is the thickness of the sample.
We perform ballistic transmission measurements for each sample by placing the scattering solution in a calibrated slab cuvette ( Hellma-Analytics) of thickness $L=\SI{1}{mm}$. The reference transmission $I_0$ is the ballistic light measured with the cuvette filled only with water (without scatterers). A picture of this samples is shown in Fig.~3a of the main text. The setup employed for measurements is illustrated in Fig.~\ref{fig:S_ls_measurement}a. The laser is firstly aligned  onto the power meter sensor without the sample. Multiple diaphragms are then placed along the beam line to filter out light with a different direction. The detector is positioned far away from the sample to determine large vertical displacements on its plane even for tiny deviations, which are then blocked by diaphragms. All the measured samples have an optical thickness  always smaller than 2 in these cuvettes, assuring that the contribution of scattered light, which inevitably reaches the detector, is always negligible compared to the ballistic component~\cite{zaccanti2003measurements}. By linearizing Eq.~(\ref{eq:BeerLambert_law}) as $\ln (I_B/I_0)=-(L/\ell'_s)\mathrm{C_f}$ the parameter $\ell'_s$ is retrieved through a linear fit of the points and the scattering mean free paths $\ell_s$ for all samples are obtained by using Eq.~(\ref{eq:ls_scaling}). In Fig.~\ref{fig:S_ls_measurement}b and Fig.~\ref{fig:S_ls_measurement}c we report data and fits of the characterization procedure for the two sample sets used in the experiment. Measured $\ell_s$ are reported in Tab.~\ref{tb:characterization_samples_poly350} and in Tab.~\ref{tb:characterization_samples_poly100}. Errors on $\ell_s$ are derived from the interval of $95\%$ statistical confidence returned by the fit on the parameters.\\ \\
\textit{Scattering anisotropy}.
The scattering diagram of dielectric spherical particles is completely described by Mie theory, once the particle radius and its refractive index are known~\cite{VanDeHulst1957light}. We have calculated $g=\langle \cos\theta \rangle$ by using the refractive index of polystyrene $n=1.5983$ at $\lambda=$\SI{532}{nm} and particles' radii are obtained by the measurement of the diffusion constant.

\begin{table}
\centering
\begin{tabular}{c c}
Polybeads \SI{350}{nm} & Polybeads \SI{100}{nm} \\
\hline
 $D= (1.30 \pm 0.08)\times 10^{-3}$ $\mu$m$^2/$ms  & $D= (4.50 \pm 0.06)\times 10^{-3}$ $\mu$m$^2/$ms \\
 $r=$\SI{181 \pm 11}{nm} & $r=$\SI{51.9 \pm 0.7}{nm}\\
 $g=0.77$ & $g=0.11$\\
\end{tabular}
\caption{Measured values of the particles diffusion constant $D$, radius $r$ and scattering anisotropy $g$ for the two colloidal solutions}
\label{tb:characterization_sigle_particle}
\end{table}
\begin{table}
\centering
\begin{tabular}{c c}
$\ell_s \pm \sigma $ (\SI{}{\micro\meter}) & $\ell^* \pm \sigma$ (\SI{}{\micro\meter}) \\
\hline
$ 1234 \pm 30 $   & $ 5367 \pm 130 $ \\
$ 617 \pm 14  $   & $ 2684 \pm 64 $\\
$ 247.1 \pm 5.9 $ & $ 1073 \pm 25 $\\
$ 123.4 \pm 3.9 $ & $ 536 \pm 13  $\\
$ 29.6 \pm 0.7  $ & $ 129.2 \pm 3.1 $\\
\end{tabular}
\caption{Measured scattering lengths for polybeads \SI{350}{nm}}
\label{tb:characterization_samples_poly350}
\end{table}
\begin{table}
\centering
\begin{tabular}{c c}
$\ell_s \pm \sigma $ (\SI{}{\micro\meter}) & $\ell^* \pm \sigma$ (\SI{}{\micro\meter})\\
\hline
$ 432 \pm 48  $  &  $485  \pm 54 $ \\
$ 1152 \pm 130  $ & $1293 \pm 143 $\\
$ 2158 \pm 240  $ & $2425 \pm 268 $\\
$ 4316 \pm 470  $ & $4849 \pm 536 $\\
\end{tabular}
\caption{Measured scattering lengths for polybeads \SI{100}{nm}}
\label{tb:characterization_samples_poly100}
\end{table}

\subsection*{Diffusion constant of the scatterers}
The diffusion constant of monodispersed particles suspended in a liquid can be obtained by Dynamic Light Scattering (DLS) measurements when the particles' concentration is low enough to guarantee single light scattering~\cite{berne2000dynamic}. In this condition the field autocorrelation of light scattered at an angle $\theta$ is expected to decay exponentially  as
\begin{equation}
g_E(\tau,\theta)=\exp \left(-\frac{\tau}{\tau_0(\theta)}\right),
\label{eq:field_autocorr_DLS}
\end{equation}
with the decay constant $\Gamma=1/\tau_0$ given by 
\begin{equation}
\Gamma(\theta)=Dq^2 (\theta),
\label{eq:decay_constant_DLS}
\end{equation}
where $q^2$ is the scattered wave vector squared and $D$ is the  particles' diffusion constant.

\begin{figure}
   \centering
   \includegraphics[scale=1]{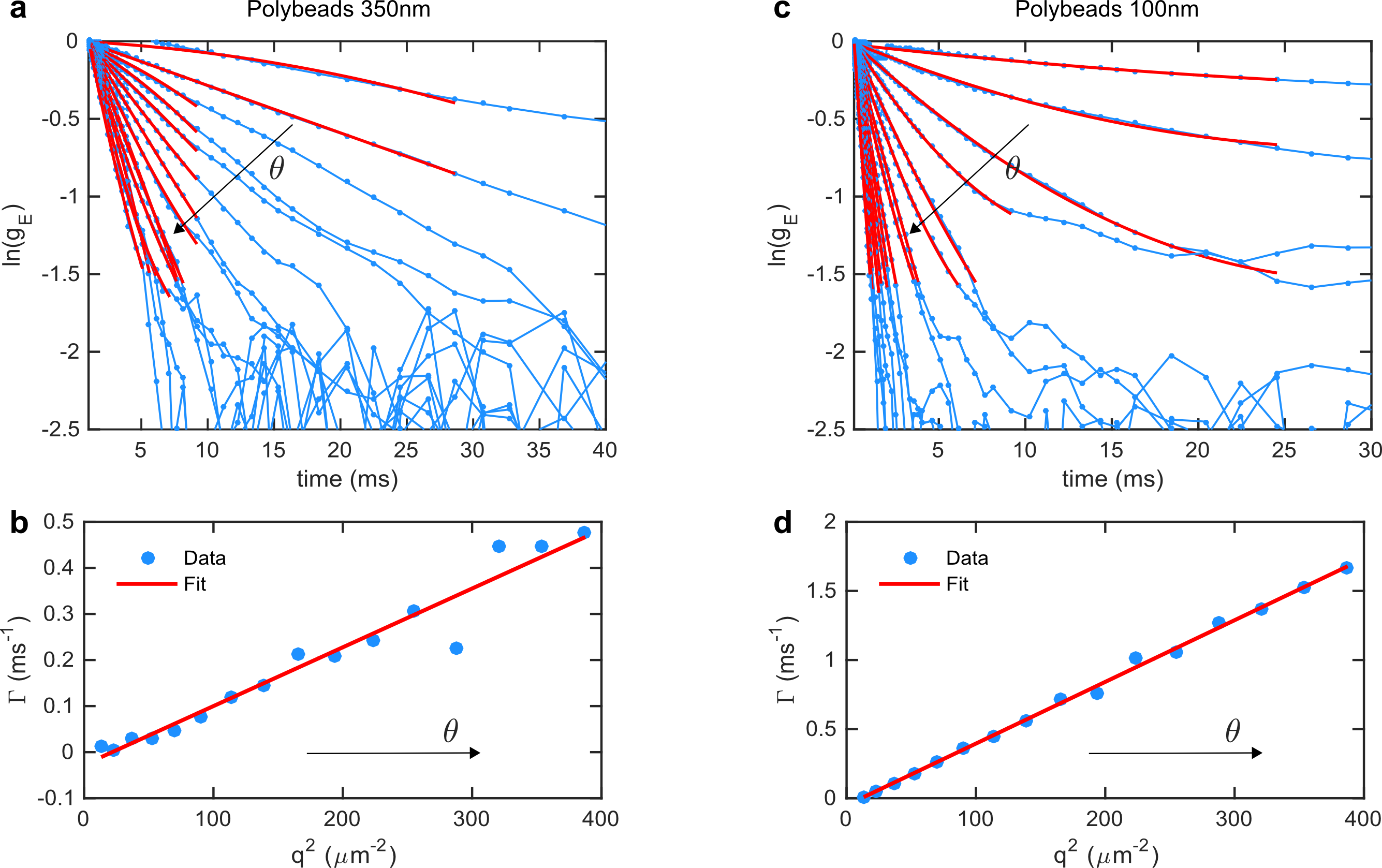}
   \caption{Procedure for the measurement diffusion constant $D$ of colloidal particles used in the experiment. \textbf{a}-\textbf{c}, Field autocorrelations measured on the DLS machine at different angles (light blue lines). Red lines correspond to the fit employed for measuring the decay constant of each curve. \textbf{b}-\textbf{d}, Data and fit based on Eq.~(\ref{eq:decay_constant_DLS}) used to measured $D$.}
   \label{fig:S_D_measurement}
\end{figure}

After having obtained sufficiently diluted  solutions we performed DLS measurements for the two types of particles by using a calibrated commercial machine (CGS-3 ALV GmbH). Measurements are reported in Fig.~\ref{fig:S_D_measurement}a and Fig.~\ref{fig:S_D_measurement}c. We performed measurements at several angles, from  $\SI{15}{\degree}$ to $\SI{90}{\degree}$ at steps of $\SI{5}{\degree}$.  When such large variation of angles is considered, residual multiply scattered light  may contribute in some cases and leads to a deviation from  the predicted exponential decay. This may be taken into account by fitting $\ln(g_E)$ with a second order polynomial and by extracting the decay constant  from the first order coefficient~\cite{hassan2014making}. Having measured the decay constant at each angle, we obtained the diffusion constant $D$ (averaged over all angles) on the base of Eq.~(\ref{eq:decay_constant_DLS}) by means of a linear fit. Data and fit are shown in  Fig.~\ref{fig:S_D_measurement}b and Fig.~\ref{fig:S_D_measurement}b, while obtained $D$ values are reported in  Tab.~\ref{tb:characterization_sigle_particle}.

We retrieved particles' radii by means of the Stokes-Einstein relation for the diffusion of spherical particles through a liquid
\begin{equation}
r=\frac{k_BT}{6\pi\eta D},
\label{eq:Stokes_Einstein}
\end{equation}
where $k_B$ is the Boltzmann's constant, $T=\SI{296.15}{K}$ ($\SI{23}{C^\degree}$) is the temperature of water and $\eta=9.35\times 10^{-4}$ Ns/m$^2$ is the viscosity of water at \SI{23}{C^\degree}. Obtained values are reported in  Tab.~\ref{tb:characterization_sigle_particle}, which are consistent with nominal values provided by the manufacturer. 

When increasing the particles' concentration the diffusion constant $D$ slightly reduces. However when the particles' volume fraction $\phi$ is kept low, the correction is normally calculated
according to the relation 
\begin{equation}
D=D_0(1-1.83\phi),
\label{eq:D_correction}
\end{equation}
which is valid up to filling fraction of \SIrange{3}{4}{\%}~(see~\cite{lemieux1999investigating} and references therein), thereby covering the concentration range of all our samples, which have a maximum volume fraction of $\phi=2.5\%$.

\section*{Path length, average time and energy velocity}

In Ref.~\cite{Pierrat:2014bp} we have derived the expression of the average length $\langle s\rangle$ under the
assumption of isotropic and uniform illumination (Eq.~(1) of main text). This expression, valid in the limit $k\ell_s$
($k$ being the wave vector and $\ell_s$ the scattering mean-free path), generalizes the initial result by Blanco and
Fournier~\cite{Blanco:2003tm} to resonant scattering. In this section, we show that this derivation can be performed
without the need of the Radiative Transfer Equation (RTE). It is based only on the hypothesis of statistical isotropy of
the specific intensity, that allows one to account for a different refractive index between the scattering medium and
the external background.

The starting point is the definition of the average time based on the analogy with the dwell time for scattering by
a single scatterer of volume $V$~\cite[Eq.~(3.37)]{LAGENDIJK-1996}
\begin{equation}
   \tau(\omega)=\frac{1}{\sigma_s(\omega)c}\int_V W(\bm{r},\omega)d^3\bm{r}
      =\frac{1}{\sigma_s(\omega)I_0}\int_V U(\bm{r},\omega)d^3\bm{r},
\end{equation}
where $\sigma_s(\omega)$ is the scattering cross section, $W(\bm{r},\omega)=U(\bm{r},\omega)/U_0(\bm{r},\omega)$ is the
normalized energy density and $I_0=cU_0$ is the incident intensity. Considering the scattering medium as a ``super-scatterer'' of volume $V$, this definition can be extended to express the averaged time spent by the wave inside the medium as
\begin{equation}
   \label{eq:average_time}
   \langle t(\omega)\rangle = \frac{1}{\phi_{\text{out}}(\omega)}\int_V U(\bm{r},\omega)d^3\bm{r},
\end{equation}
where $\phi_{\text{out}}(\omega)$ is the power scattered outside the medium (outgoing flux). Here
$\phi_{\text{out}}(\omega)$ and $U(\bm{r},\omega)$ are ensemble averaged quantities. We now introduce in the wave
formalism the specific intensity $I(\bm{r},\bm{u},\omega)$ as the Wigner transform of the field~\cite{APRESYAN-1996}. 
In
this formalism, the energy velocity is introduced by the connection between the energy density and the specific
intensity~\footnote{In terms of the specific intensity, the energy current (Poynting vector) is defined as
$\bm{J}(\bm{r},\omega)=\int_{4\pi}I(\bm{r},\bm{u},\omega)d\bm{u}$. Assuming for example a directional specific
intensity $I(\bm{r},\bm{u},\omega)=A(\bm{r},\omega)\delta(\bm{u}-\bm{u}_0)$, where $\bm{u}_0$ is a unit vector, one has
$U(\bm{r},\omega)=A(\bm{r},\omega)/v_E(\omega)$ and
$\bm{J}(\bm{r},\omega)=A(\bm{r},\omega)\bm{u}_0=U(\bm{r},\omega)v_E\bm{u}$ which is consistent with the usual expression
of the current in terms of transport velocity and density.}:
\begin{equation}
   \label{eq:energy_density}
   U(\bm{r},\omega)=\frac{1}{v_E(\omega)}\int_{4\pi}I_{\text{int}}(\bm{r},\bm{u},\omega)d\bm{u}
\end{equation}
where $I_{\text{int}}(\bm{r},\bm{u},\omega)$ is the specific intensity inside the medium, and $d\bm{u}$ means
integration over the solid angle. The outgoing flux can be written as
\begin{equation}
   \label{eq:outgoing_flux}
   \phi_{\text{out}}=\int_{\Sigma}\int_{2\pi}I_{\text{ext}}(\bm{r},\bm{u},\omega)\bm{u}\cdot\bm{n}d^2\bm{r}
\end{equation}
where $I_\text{ext}$ is the specific intensity outside the medium, $\Sigma$ is the external surface of the medium (a surface that is outside the medium but in contact with it), and
$\bm{n}$ is the outward normal. Only $\bm{u}\cdot\bm{n}>0$ is involved in the integral.

We now assume that the radiation field is statistically homogeneous and isotropic (\ie independent on $\bm{r}$ and
$\bm{u}$) inside and outside the medium~\cite{YABLONOVITCH-1982}. 
This occurs for example when the scattering medium is statistically homogeneous
and isotropic, and immersed in an incoming radiation that is itself homogeneous and isotropic (blackbody radiation is an
example of such a field). It is important to note that this assumption is an approximation when there is an index
mismatch between the internal and the external medium especially because of angular redistribution of energy at the
interface. However, this approximation is accurate when scattering occurs inside the medium.  Under these hypotheses,
the energy density and specific intensity in Eqs.~(\ref{eq:energy_density})-(\ref{eq:outgoing_flux}) write
\begin{equation}
   U(\omega)=\frac{4\pi}{v_E}I_{\text{int}}(\omega)
   \quad\text{and}\quad
   \phi_{\text{out}}(\omega)=\pi\Sigma I_{\text{ext}}(\omega).
\end{equation}
Inserting these two expressions in Eq.~(\ref{eq:average_time}) leads to
\begin{equation}
   \label{eq:average_length}
   \langle s_{\text{theo}}(\omega)\rangle=\langle t(\omega)\rangle v_E(\omega) = \frac{4V}{\Sigma}\frac{I_{\text{int}}(\omega)}{I_{\text{ext}}(\omega)}.
\end{equation}

For index-matched media (no internal reflection), the specific intensity is continuous and the last factor is unity, so
that we recover the result in Eq.~(1) of the main text. In the case of index mismatch, we need to evaluate the last
factor.

\begin{figure}
   \centering
   \includegraphics[width=0.4\linewidth]{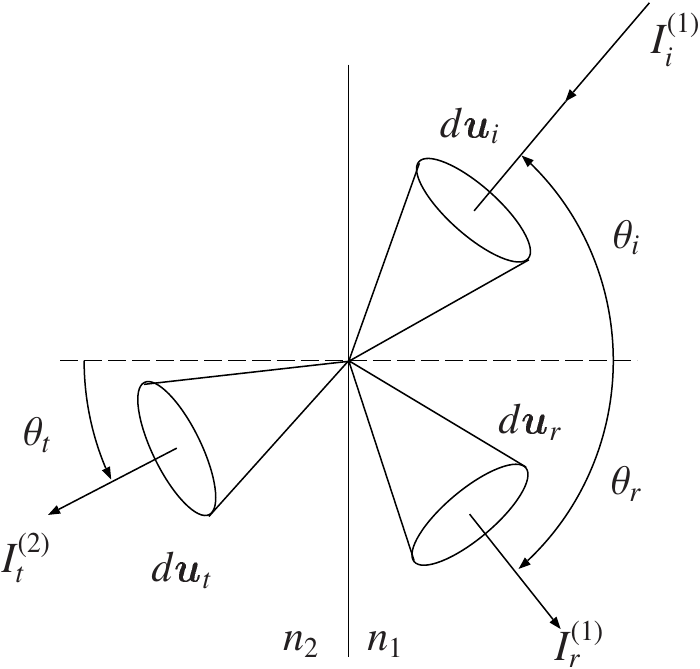}
   \caption{Reflection and transmission of an incident specific intensity at an interface between two media with
   different refractive index.}
   \label{fig:interface}
\end{figure}

With reference to Fig.~\ref{fig:interface}, let us consider two media $1$ and $2$, with refractive indices $n_1$ and
$n_2$ and subject to isotropic radiation on both sides (only radiation incident from medium $1$ is shown on the figure,
but a symmetric radiation incident from medium $2$ is assumed).

Denoting by $I_i^{(1)}$ the specific intensity incident from medium $1$, $I_r^{(1)}$ the reflected specific intensity in
medium $1$ and $I_t^{(2)}$ the transmitted specific intensity in medium $2$, energy conservation at the interface reads
\begin{equation}
   I_i^{(1)}\cos\theta_i d\bm{u}_i=I_r^{(1)}\cos\theta_r d\bm{u}_r+I_t^{(2)}\cos\theta_t d\bm{u}_t,
\end{equation}
and using Snell's law leads to
\begin{equation}
   \label{eq:snell1}
   I_i^{(1)}=I_r^{(1)}+I_t^{(2)}\frac{n_1^2}{n_2^2}.
\end{equation}
Using symmetric notations for a specific intensity incident from medium $2$, we obtain
\begin{equation}
   \label{eq:snell2}
   I_i^{(2)}=I_r^{(2)}+I_t^{(1)}\frac{n_2^2}{n_1^2}.
\end{equation}
Since the radiation on both sides of the interface is assumed to be isotropic, we also have
\begin{equation}
   \label{eq:flux1}
   I_i^{(1)}=I_r^{(1)}+I_t^{(1)}
\end{equation}
and
\begin{equation}
   \label{eq:flux2}
   I_i^{(2)}=I_r^{(2)}+I_t^{(2)}
\end{equation}
which can be established, for example, by writing that the flux through a surface in each medium has to vanish. From
Eqs.~(\ref{eq:snell1}-\ref{eq:flux2}), it is easy to show that $I_{i,r,t}^{(2)}=(n_2^2/n_1^2)I_{i,r,t}^{(1)}$. In other
words, assuming isotropic radiation on both sides, the specific intensity in medium $2$ is greater than the specific
intensity in medium $1$ by a factor $n_2^2/n_1^2$. 

Coming back to Eq.~(\ref{eq:average_length}), we therefore end up with the following expression of the averaged path
length:
\begin{equation}
   \langle s_{\text{theo}}(\omega)\rangle=\langle t(\omega)\rangle v_E(\omega) = \frac{4V}{\Sigma}\frac{n_2^2}{n_1^2}.
\end{equation}
This result can also be recovered using a Density of Optical States formalism~\cite{MELCHARD-2015}.
\begin{figure}
   \centering
   \includegraphics[width=0.4\linewidth]{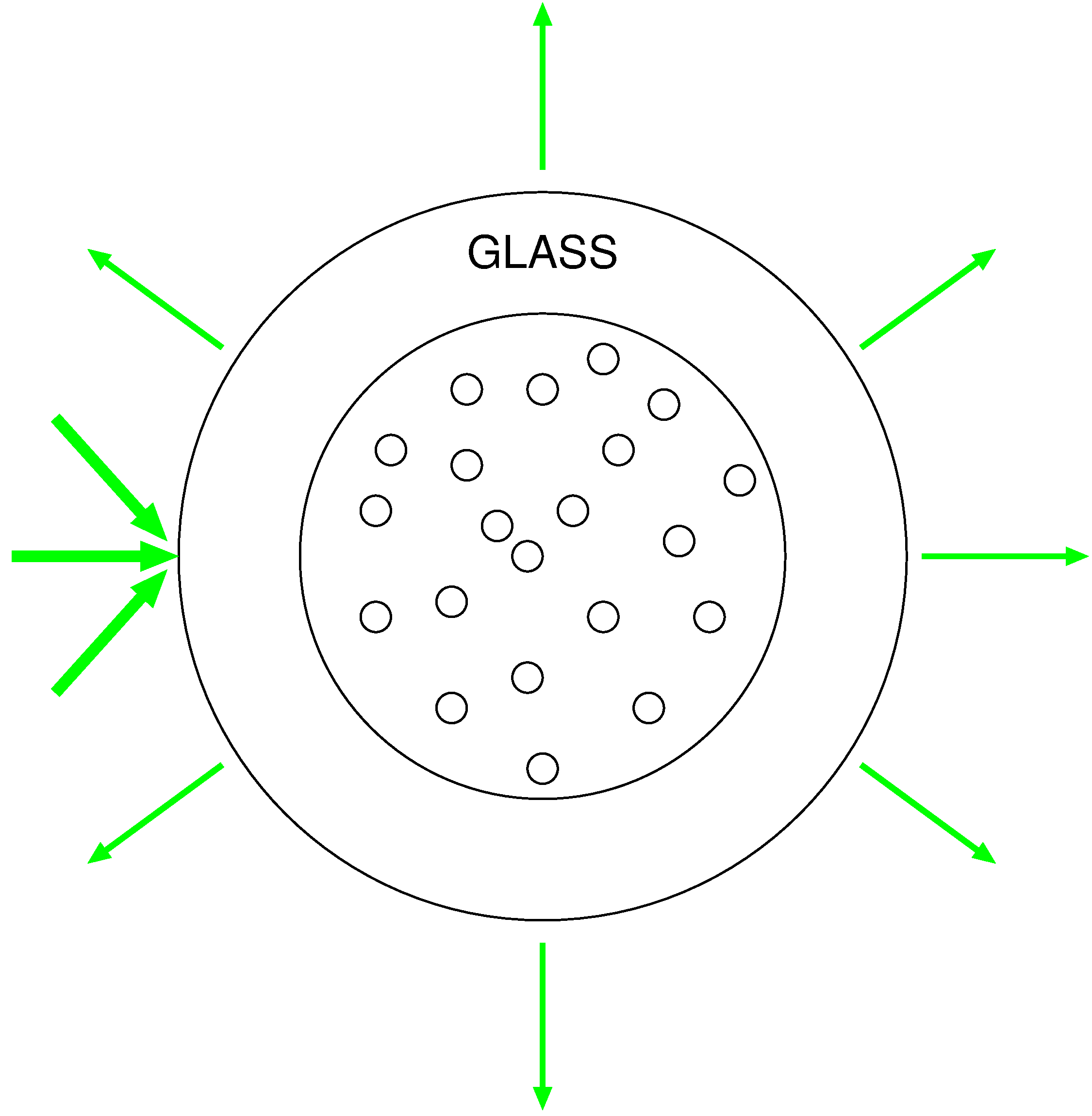}
   \caption{Schematic view of the system considered in the Monte Carlo simulation. To mimic the experiment, the
   scattering region is a cylinder of radius $R=\SI{4.27}{mm}$ filled with water (refractive index $n=\num{1.33}$) and
   polystyrene spherical particles of radius $a=\SI{0.175}{\micro\meter}$ (refractive index $n_p=\num{1.6}$). The scattering region
   is contained in a glass cell delimited also by a cylinder of radius $R'=\SI{4.87}{mm}$ (refractive index
   $n_s=\num{1.56}$). The entire system is illuminated by a homogeneous and isotropic beam ($\lambda=\SI{532}{nm}$) and all
   photons escaping the system are collected.}
   \label{fig:S_mc}
\end{figure}

\section*{Monte Carlo simulations}
 Monte Carlo simulations have been performed to evaluate
\begin{itemize}
   \item the accuracy of the Diffusing Wave Spectroscopy  method to measure the average path length;
   \item and the effect of the glass cell on the average path length (flatness and shift).
\end{itemize}

For that purpose, we have considered a geometry as close as possible to the experiment (see Fig.~\ref{fig:S_mc}).  The
scattering parameters [\ie scattering mean free path $\ell_s$ and phase function $p(\bm{u}\cdot\bm{u}')$] are computed
using the Mie theory and are averaged over the polarization state. As in the experiment, the concentration of beads is
tuned to vary the transport mean free path $\ell^*$.

The average path length is computed using two different methods: The first is the theoretical one where $\langle
s\rangle$ is simply given by the average of all path lengths simulated by the Monte Carlo process. The second one mimics
the experimental procedure based on $g_E(\tau)$. To be as close as possible to the experiment, we have to compute
$g_E(\tau)$ with a formalism beyond the standard DWS, which is valid in the diffusive regime only. The starting point
is a transport equation for the field autocorrelation function
~\cite{ACKERSON-1992,DOUGHERTY-1994,PIERRAT-2008-1}, valid
for all transport regimes from ballistic to diffusive:
\begin{equation}
   \left[\bm{u}\cdot\bm{\nabla}_{\bm{r}}+\frac{1}{\ell_s}\right]g_E(\bm{r},\bm{u},\tau)
      =\frac{1}{4\pi\ell_s}\int p(\bm{u}\cdot\bm{u}')g(k|\bm{u}-\bm{u}'|\tau)g_E(\bm{r},\bm{u}',\tau)d\bm{u}'
\end{equation}
where
\begin{equation}
   g(k|\bm{u}-\bm{u}'|\tau)=\exp\left[-|\bm{u}-\bm{u}'|^2\frac{\tau}{\tau_0}\right].
\end{equation}
This transport equation is very similar to the classical one used to describe the evolution of the steady-state specific
intensity $I(\bm{r},\bm{u})$. Thus the Monte Carlo scheme used to solve it numerically is exactly the same except that
we have to multiply by $g(k|\bm{u}-\bm{u}'|\tau)$ at each scattering event. Denoting by $\Theta$ the scattering angle of
a given scattering event, we have
\begin{equation}
   g_E(\tau)=\frac{1}{N}\sum_{i=1}^N\prod_{j=1}^{n_i}
      \exp\left[-2\left(1-\cos\Theta_{i,j}\right)\frac{\tau}{\tau_0}\right]
\end{equation}
where $N$ is the number of Monte Carlo shots performed and $n_i$ the number of scattering events for shot number $i$.
Finally, applying Eq.~(3) of the main text leads to
\begin{equation}
   \langle s_{\text{DWS}}\rangle=-\left.\frac{dg_E}{d\tau}\right\vert _{\tau=0} \frac{\ell^*\lambda_0^2}{8\pi^2n^2D}
      =\frac{1}{N}\sum_{i=1}^N\sum_{j=1}^{n_i}\ell^*\left(1-\cos\Theta_{i,j}\right)
\end{equation}
which is the formula used in the Monte Carlo simulation to have an estimate of the average path length as it is measured
experimentally. In the diffusive limit, $n_i\gg 1$ such that $\cos\Theta_{i,j}\sim g$ we have
\begin{equation}
   \langle s_{\text{DWS}}\rangle\sim \frac{1}{N}\sum_{i=1}^N n_i\ell_s
      \sim \frac{1}{N}\sum_{i=1}^N s_i \sim \langle s_{\text{in}}\rangle
\end{equation}
where $\langle s_{\text{in}}\rangle$ corresponds to the average path length inside the scattering region.

\begin{figure}
   \centering
   \includegraphics[width=0.5\linewidth]{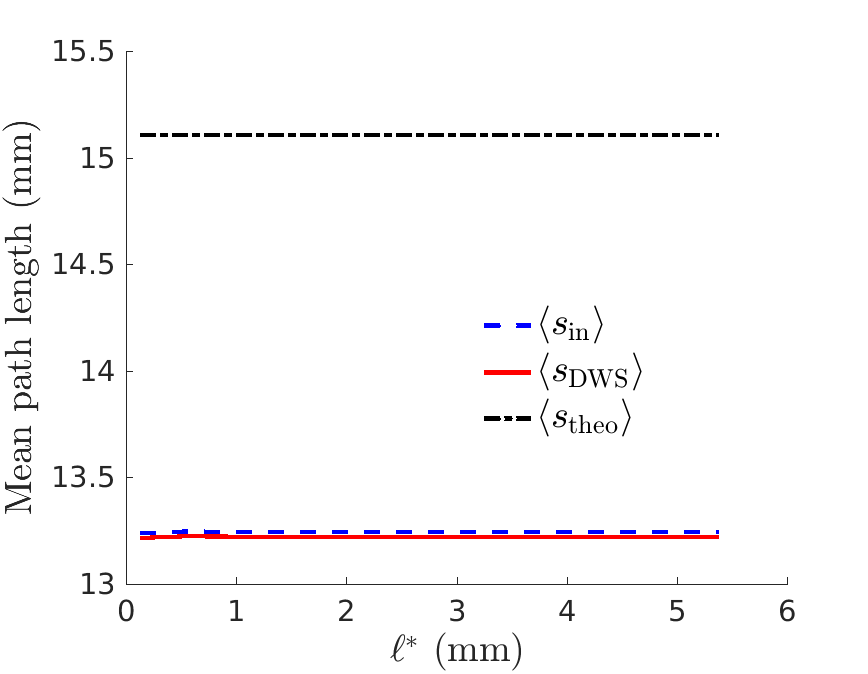}
   \caption{Average path length as a function of the transport mean free path. The theoretical path length
   $\langle s_{\text{theo}}\rangle$ is compared to the one derived from the computation of the field autocorrelation
   function $\langle s_{\text{DWS}}\rangle$ and to the average path length inside the scattering region $\langle
   s_{\text{in}}\rangle$. The discrepancy between $\langle s_\text{in}\rangle$ and $\langle s_{\text{theo}}\rangle$ is due to
   the glass cell (role of the second interface of radius $R'$). The good agreement between $\langle
   s_{\text{in}}\rangle$ and $\langle s_{\text{DWS}}\rangle$ indicates that the use of the DWS method is
   relevant to measure the average path length inside the scattering region. As for $\langle s_{theo}\rangle$, we note also
   that $\langle s_{\text{in}}\rangle$ is independent of the transport regime.}
   \label{fig:S_average_length} 
\end{figure}

Numerical results are shown in Fig.~\ref{fig:S_average_length}. Several predictions are compared in this figure:
\begin{itemize}
   \item The theoretical path length $\langle s_{\text{theo}}\rangle$. In the case of the cylinder geometry considered here,
   $4V/\Sigma=2R$ and $n_2^2/n_1^2\sim n^2$ (the system being diluted, the effective refractive index is close to the
   refractive index of water $n$). This leads to
   \begin{equation}
      \langle s_{\text{theo}}\rangle=2Rn^2.
   \end{equation}
   \item The average path length computed from the field autocorrelation function $\langle s_{\text{DWS}}\rangle$.
   \item The average path length restricted to the scattering region $\langle s_{\text{in}}\rangle$.
\end{itemize}

\bibliographystyle{unsrt}
\bibliography{S_bibliography.bib}